\documentclass[a4paper,11pt]{article}
\pdfoutput=1 

\usepackage{jcappub} 

\usepackage[T1]{fontenc} 
\usepackage{amsmath}
\usepackage{caption}
\usepackage{hyperref}
\usepackage[utf8]{inputenc}
\usepackage{bm}
\usepackage{wrapfig}
\hypersetup{
    colorlinks=true,
    linkcolor=blue,   
    urlcolor=blue,
}

\title{\boldmath Trading kinetic energy: How late kinetic decoupling of dark matter changes $N_{\textrm{eff}}$}

\author{James A.~D.~Diacoumis}
\author{and Yvonne Y.~Y.~Wong}

\affiliation{School of Physics, The University of New South Wales, Sydney NSW 2052, Australia}

\emailAdd{j.diacoumis@unsw.edu.au}
\emailAdd{yvonne.y.wong@unsw.edu.au}

\abstract{Elastic scattering between dark matter particles and a relativistic species such as photons or neutrinos leads to a transfer of energy from the latter due to their intrinsically different temperature scaling relations. In this work, we point out that this siphoning of energy from the radiation bath manifests as a change in the effective number of neutrinos \(N_{\textrm{eff}}\), and compute the expected shift \(\Delta N_{\textrm{eff}}\) for dark matter--photon and dark matter--neutrino elastic scattering as a function of the dark matter mass~$m_\psi$ and scattering cross section~$\sigma_{\psi-X}$.
For $(m_\psi,\sigma_{\psi-X})$-parameter regions already explored by nonlinear probes such as the Lyman-$\alpha$ forest through collisional and/or free-streaming damping, we find shifts of $|\Delta N_{\rm eff}| \simeq O(10^{-2})$, which may be within the reach of the proposed CMB-S4 experiment.  For most of the as-yet-unexplored parameter space, however, we expect $|\Delta N_{\rm eff}| \lesssim O(10^{-3})$.
An ideal 21~cm tomography survey of the dark ages  limited only by cosmic variance is potentially sensitive to $|\Delta N_{\rm eff}| \simeq O(10^{-6})$, in which case dark matter masses up to \(m_{\psi} \sim 10~\textrm{MeV}\) may be probed via their effect on $N_{\rm eff}$.}

\begin{document}
\maketitle
\flushbottom

\section{Introduction}

The standard $\Lambda$CDM paradigm of modern cosmology has enjoyed remarkable success at explaining the structure of the universe on the largest observable scales.  In this worldview, some 26\% of the universe's present-day energy density is in the form of cold dark matter, 69\% dark energy, and baryonic matter makes up the remaining 5\%. The advent of cosmic microwave background (CMB) anisotropies measurements in the past two decades by, e.g., the WMAP~\cite{Spergel:2003cb}~and Planck~\cite{Planck, Aghanim:2018eyx}~missions, has even seen these numbers determined to better than 1\% precision, with a ten-fold improvement anticipated from the forthcoming CMB-S4 experiment~\cite{Abazajian:2016yjj}.
Cosmology as a precision science has truly come of age.

Notwithstanding the success of $\Lambda$CDM, the remarkable precision with which we are now able to infer the properties of the universe from observations also invites us to question and test the assumptions underpinning the paradigm, through which to search for new physics.  One such assumption is the non-photon radiation energy density, conventionally parameterised by the effective number of neutrinos $N_{\rm eff}$. Assuming standard model physics, this number is predicted to be  $N_{\rm eff}^{\rm SM}=3.046$~\cite{Mangano:2005cc, deSalas:2016ztq, deSalas:2017wtt}, accounting for the energy density in three ideal gases of standard model neutrinos, plus corrections from neutrino energy transport and finite-temperature quantum electrodynamics. 

There are however many ways in which $N_{\rm eff}$ can be changed by new physics, including the thermalisation of a new light particle~\cite{Ackerman:mha, Abazajian:2012ys, Boehm:2012gr}, decay of a heavy particle~\cite{Hasenkamp:2012ii, Hasenkamp:2014hma}, 
late-time conversion~\cite{Bringmann:2018jpr},  
large lepton asymmetries~\cite{Mangano:2011ip,Oldengott:2017tzj}, and non-standard reheating after inflation~\cite{Kobayashi:2011hp}.
Current cosmological observations prefer $N_{\rm eff} = 2.99 \pm 0.17~$(68\% C.L.)~\cite{Aghanim:2018eyx}, consistent with the standard-model value, and hence placing strong constraints on the viable new physics model parameter spaces.  In the near future,  conservative configurations of CMB-S4 can be expected to improve the $1 \sigma$ uncertainty on $N_{\rm eff}$ to $\pm 0.02 \to 0.03$, potentially probing an even wider range of new physics.

In this connection, it is interesting to note that all of the aforementioned new physics that can change $N_{\rm eff}$ do so through an increase or decrease in the {\it number density} of relativistic particles relative to the photon number density. In this work, we point out that a non-standard $N_{\rm eff}$ can also arise without invoking number-changing processes.  Interactions that modify {\it only} the energy content, in particular, elastic scattering with a non-relativistic particle species, can transfer energy out of the photon or neutrino population, leading to an increase or decrease in the neutrino energy density relative to its photon counterpart, thereby altering~$N_{\rm eff}$.

The particular scenario we have in mind is that of Late Kinetic Decoupling (LKD)~\cite{InteractingDM, InteractingDM2, LKD, LKD2, LKD3}, whereby elastic scattering between relativistic neutrinos or photons and non-relativistic cold dark matter keeps the participating particle species in kinetic equilibrium until well after the big bang nucleosynthesis epoch in time.  The scenario is especially interesting from the point of view of  structure formation on small length scales: when the scattering rate per dark matter particle exceeds the Hubble expansion rate, the interaction enables the dark matter density perturbations to closely track  relativistic perturbations  on subhorizon scales during radiation domination, leading to a strong suppression of growth on cluster and galactic scales that may provide a parsimonious and theoretically appealing solution to the small-scale problems of cold dark matter~\cite{MissingSat1,MissingSat2,TBTF1,TBTF2,CuspCore1,CuspCore2,CuspCore3}.  The phenomenologies of LKD has also been considered in the context of CMB anisotropies~\cite{Wilknu, Wilkphoton, Diacoumis:2018ezi} and spectral distortions~\cite{Ali-Haimoud:2015pwa, Diacoumis:2017hff}.

The paper is organised as follows.   We describe in section~\ref{Sec:Def} how elastic scattering between non-relativistic dark matter and radiation can alter $N_{\rm eff}$, and introduce in section~\ref{sec:formalism} a formalism based on equilibrium thermodynamics with which to compute these changes.  We apply this formalism to the specific scenarios of dark matter--photon and dark matter--neutrino coupling respectively in  sections~\ref{Sec:Nu} and \ref{Sec:Phot}, and compute in each case \(N_{\textrm{eff}}\)  as a function of the elastic scattering cross section and dark matter particle mass.  In section~\ref{Sec:Forecast} we discuss the impact of our results on future experiments, and perform a parameter sensitivity forecast for an idealised 21~cm tomography survey limited only by cosmic variance.  Our conclusions are presented in section \ref{Sec:Conclusion}.


\section{The scenario}
 \label{Sec:Def}
 
The effective number of neutrinos parameter $N_{\rm eff}$  formally parameterises the neutrino energy density $\rho_\nu$ in the era between electron--positron annihilation (at temperature $T \sim m_e$, where $m_e$ is the electron mass) and when any one of the neutrinos transitions from a relativistic to non-relativistic particle species (at $T \sim m_\nu$).  Expressed in terms of  the energy density in photons $\rho_\gamma$~\cite{NuCosmology},
\begin{equation} \label{eq:Defn}
\rho_{\nu} \equiv N_{\textrm{eff}}\frac{7}{8}\left(\frac{T_{\nu}}{T_{\gamma}}\right)^{4} \rho_\gamma,
\end{equation}
where the factor \(7/8\) accounts for the Fermi--Dirac spin statistics of neutrinos, and the temperature ratio \(T_{\nu}/T_{\gamma} = \left(4/11\right)^{1/3}\) originates from electron--positron annihilation which raises the temperature of the photon bath.  

The standard model of particle physics predicts the effective number of neutrinos to be \(N_{\textrm{eff}}^{\textrm{SM}} = 3.046\)~\cite{Mangano:2005cc}, representing the energy density in three ideal gases of standard model neutrinos, each with two internal degrees of freedom. The small percent-level correction arises from (i) neutrino energy transport, which enables some of the energy from the $e^+ e^-$ annihilation to heat up the neutrino population, 
and (ii) finite-temperature quantum electrodynamics, which alters the equation of state of the photon--electron--positron plasma.
Any deviation from the standard scenario can be parametrised by \(N_{\textrm{eff}} = N_{\textrm{eff}}^{\textrm{SM}} + \Delta N_{\textrm{eff}}\). 
In terms of small changes in the neutrino or photon energy density, it is immediately clear from equation~(\ref{eq:Defn}) that 
\begin{equation}
\begin{aligned} 
\label{eq:deltaneff}
 &\frac{\Delta N_{\textrm{eff}}}{N^{\textrm{SM}}_{\textrm{eff}}} \simeq \frac{\Delta \rho_{\nu}}{\rho_{\nu}^{\textrm{SM}}}, \\
& \frac{\Delta N_{\textrm{eff}}}{N^{\textrm{SM}}_{\textrm{eff}}} \simeq - \frac{\Delta \rho_{\gamma}}{\rho_{\gamma}^{\textrm{SM}}} ,
\end{aligned}
\end{equation}
where \(\Delta \rho_{\nu} \ll \rho_{\nu}\) and \(\Delta \rho_{\gamma} \ll \rho_{\gamma}\) are understood.

When a non-relativistic cold dark matter is allowed to couple elastically to either the neutrinos or the photons in the era where the definition~(\ref{eq:Defn}) holds, kinetic energy exchange between the dark matter and relativistic sectors can siphon energy out of the latter simply as a consequence of equipartition (when kinetic equilibrium holds) and the fact that kinetic energy evolves with the scale factor~$a$ 
 at different rates for fully relativistic ($p \propto a^{-1}$) and fully non-relativistic ($p^2/2m \propto a^{-2}$) particle species.
 If, for example, the dark matter couples to (relativistic) neutrinos, then one would expect the neutrino energy density~$\rho_\nu$ to decrease with an increasing $a$ a little ``faster'' than the standard  $\rho_\nu \propto a^{-4}$.  Over time a non-zero and negative $\Delta \rho_\nu$ will develop, leading to a negative change in $N_{\rm eff}$.  The same is expected to occur in the photon sector if the dark matter were to couple to photons, except of course in this case the change in $N_{\rm eff}$ would be positive according to equation~(\ref{eq:deltaneff}).

Our goal in the next section is to write down a general formalism based on equilibrium thermodynamics that will enable us to track the change in the relativistic energy density as a consequence of such couplings.  Application of this formalism to specific LKD scenarios will be discussed in sections~\ref{Sec:Nu} and~\ref{Sec:Phot}.

 
 \section{Equilibrium  thermodynamics formalism}
 \label{sec:formalism}
 
We focus on the epoch immediately after neutrino decoupling at $T\sim 1$~MeV.  Electron--positron annihilation follows shortly after at $T \sim 0.5$~MeV, so that only photons~$\gamma$, neutrinos~$\nu$, and dark matter~$\psi$ remain in significant abundance. The dark matter $\psi$ can exchange energy with either the neutrino or the photon heat bath via elastic scattering but has no number-changing processes. We assume that the scattering rate is sufficiently high so that kinetic equilibrium between the dark matter and the radiation is always maintained, and all relevant particle phase space densities can be quantified by equilibrium distributions characterised by the species' temperature $T_i$ and chemical potential $\mu_i$.

With these assumptions in mind, we can immediately write down the following quantitative descriptions:
\begin{enumerate}
\item Following from the zeroth law of thermodynamics, the temperatures of the dark matter  and the radiation to which it couples must be the same, i.e., 
\begin{equation}
\label{eq:sametemp}
T_\psi = T_X,
\end{equation} 
where $X$ may be $\nu$ or $\gamma$ according to the case in hand.

\item The comoving number densities of dark matter and neutrinos are individually  conserved, 
\begin{equation}
\label{eq:numbercon}
 \textrm{d}\left(n_i a^{3} \right) = 0,
\end{equation}
where $i = \psi, \nu$.  The photon number density can change via double Compton scattering and Bremsstrahlung in the presence of trace amounts of free electrons and nuclei down to a temperature of $T \sim O(500)$~eV~\cite{Chluba:2015bqa}; at lower temperatures, however, equation~(\ref{eq:numbercon}) applies also to $i= \gamma$.

\item Since equilibrium phase space distributions always apply to coupled particle species, which implies the  coupled dark matter--radiation system is always in a maximal entropy state, the total comoving entropy of the coupled system  must be conserved, i.e.,   
\begin{equation}
\label{eq:entropycon}
\textrm{d}\left(\sum_i s_i a^{3} \right) = 0,
\end{equation}
 where 
\begin{equation}
\label{eq:entropy}
s_{i}(a) \equiv \frac{\rho_{i} + P_{i} - \mu_{i}n_{i}}{T_i} \equiv \frac{\rho_{i} + P_{i}}{T_i} - \xi_{i} n_{i}
\end{equation}
is the entropy density of particle species \(i\) in the coupled system, with \(\rho_{i}, n_{i},  P_{i}\) and \(\mu_{i}\) denoting respectively the species' energy density, number density, pressure, and chemical potential, and  we have defined the degeneracy parameter \(\xi_{i} \equiv \mu_{i}/T_i\).
\end{enumerate}
 Our overall strategy then is to use the conservation laws~(\ref{eq:numbercon}) and~(\ref{eq:entropycon}) to solve for the common temperature $T_\psi=T_X$ and the chemical potentials~$\mu_i$ as functions of the scale factor~$a$ between neutrino decoupling and dark matter kinetic decoupling, from which to determine the shift in the relevant radiation energy density.  We detail below the specific thermodynamics expressions employed for each particle species.


\subsection{Dark matter}

In general, the equilibrium dark matter phase space density can be described by either the Fermi--Dirac or Bose--Einstein distribution, depending on the DM particle's spin. Furthermore, we expect the dark matter population to acquire a non-zero chemical potential, denoted by the degeneracy parameter~$\xi_\psi$, because by definition the dark matter can only exchange energy with the radiation heat bath but has no number-changing processes.

The present-day photon number density has been measured by COBE FIRAS  to be $n_\gamma=410.7 \pm 0.3~{\rm cm}^{-3}$~\cite{Fixsen:1996nj,Tanabashi:2018oca}; the standard model predicts a similar number for three families of light neutrinos, $n_\nu \simeq 336~{\rm cm}^{-3}$~\cite{NuCosmology}.
 On the other hand, current observations prefer a dark matter energy density  of  $\rho_\psi = \Omega_\psi \rho_{\rm crit} \simeq 1.25 \times 10^{-6}~{\rm GeV} \ {\rm cm}^{-3}$ within vanilla $\Lambda$CDM~\cite{Aghanim:2018eyx}.
For a minimum DM particle mass of  \(1 \, \textrm{keV}\), these numbers imply that the dark matter number density must be at least  \(O(100)\) times smaller than the radiation number density. Demanding that the dark matter and radiation populations share the same temperature while simultaneously respecting such a huge number density disparity immediately tells us that \(\xi_{\psi}\) must be negative and \(\textrm{exp}\left(|\xi_{\psi}|\right) \gg 1\) when the species are coupled.  

It then follows that,  irrespective of whether the dark matter particle is fermionic or bosonic, we may approximate its equilibrium phase space distribution as 
\begin{equation}
f\left(\epsilon, T_\psi, \xi_{\psi}, a\right) \simeq \textrm{exp}\left(-\frac{\epsilon}{a T_\psi} - |\xi_{\psi}|\right),
\end{equation}
where $\epsilon \equiv a E$ is the comoving single-particle energy.  Integrating $f_\psi$ over momentum we find the number density
\begin{equation}
n_{\psi} (a) = \frac{g_{\psi} T_\psi^{3}}{2 \pi^{2}} \,  \textrm{e}^{-|\xi_{\psi}|} x^{2} K_{2}(x),
\end{equation}
where \(x \equiv m_{\psi}/T_\psi\), $m_\psi$ is the DM mass, $g_\psi$ the DM internal degrees of freedom, and $K_n(x)$ denotes the $n$th order modified Bessel function of the second kind.  From this we can recast
the conservation of comoving DM number density~(\ref{eq:numbercon}) as an equation of motion for the DM chemical potential \(\xi_{\psi}\),
\begin{equation} \label{eq:DMxi}
\frac{|\xi_{\psi}|^{'}}{3} = \frac{1}{a} + \left[1+\frac{x}{3}\frac{K_{1}(x)}{K_{2}(x)}\right]\frac{T_\psi^{'}}{T_\psi} ,
\end{equation}
where the dash denotes differentiation with respect to the scale factor \(a\),
 and we remind the reader that $T_\psi$ tracks the temperature of the radiation to which it  couples by equation~(\ref{eq:sametemp}).

Likewise, following equation~(\ref{eq:entropy}) we can write down an expression for the dark matter entropy density,
\begin{align} \label{eq:DMentropy}
\begin{split}
s_{\psi}(a) &= \frac{g_{\psi} T_\psi^{3}}{2 \pi^{2}} \, \textrm{e}^{-|\xi_{\psi}|} \left[\left(4 + |\xi_{\psi}|\right)x^{2}K_{2}(x) +x^{3} K_{1}(x) \right] \\
&= n_{\psi}(a) \left[4 + |\xi_{\psi}| + x \frac{K_{1}(x)}{K_{2}(x)}\right],
\end{split}
\end{align}
which will be used in section~\ref{sec:master} when we demand that the comoving entropy in DM and radiation be jointly conserved.


\subsection{Radiation}
When in kinetic equilibrium and at temperatures \(T_{\nu} \gg m_{\nu}\), the neutrino and photon phase space distributions are described respectively by the relativistic Fermi--Dirac  and the relativistic Bose--Einstein  distribution,
\begin{equation}
f_{X}(q, T_X, \xi_{X}, a) = \frac{1}{\textrm{exp}\left(\frac{q}{a T_X}-\xi_{X}\right)\pm1},
\end{equation}
where $q \equiv a p$ is the comoving single-particle momentum, and the ``$+/-$'' sign applies to $X=\nu$ and $X=\gamma$ respectively.%
\footnote{In the following, whenever a choice of signs is displayed, the upper option always corresponds to the case of Fermi--Dirac statistics and the lower option to Bose--Einstein statistics.}
We expect the degeneracy parameter $\xi_X$ to be vanishingly small prior to neutrino decoupling, but to develop in general to a sizeable value afterwards if the radiation scatters elastically with the dark matter. 

The corresponding number, energy, and entropy densities are given by
\begin{align} 
\label{eq:nunum}
n_X(a) &= \mp \frac{g_XT_X^{3}}{\pi^{2}} \, \textrm{Li}_{3}\left(\mp \textrm{e}^{\xi_X}\right), \\
\rho_X(a) &=\mp \frac{3 g_X T_X^{4}}{2 \pi^{2}} \, \textrm{Li}_{4}\left(\mp \textrm{e}^{\xi_X}\right), \label{energydensitynu} \\
s_X(a) &= \mp \frac{4 g_XT_X^{3}}{\pi^{2}} \, \textrm{Li}_{4}\left(\mp\textrm{e}^{\xi_X}\right) - \xi_X n_X(a) \nonumber \\
&= n_X(a) \, \left[ 4 \frac{\textrm{Li}_{4}\left(\mp \textrm{e}^{\xi{_X}}\right)}{\textrm{Li}_{3}\left(\mp\textrm{e}^{\xi_X}\right)}-\xi_X\right], \label{eq:nuentropy}
\end{align}
in which the function \(\textrm{Li}_{n}(z)\) denotes the polylogarithm of \(n\)th order, and we note that the argument $ \textrm{e}^{\xi_X}$ is but the fugacity of the gas.
Where conservation of the comoving radiation number density applies, we can use the number density~(\ref{eq:nunum}) to rewrite the conservation law~(\ref{eq:numbercon}) as
\begin{equation} 
\label{eq:nuxi}
 \frac{\xi_{X}^{'}}{3} = -  \frac{\textrm{Li}_{3}\left(\mp \textrm{e}^{\xi_{X}}\right)}{\textrm{Li}_{2}\left( \mp \textrm{e}^{\xi_{X}}\right)}  \left[\frac{1}{a} + \frac{T_X^{'}}{T_X} \right],
\end{equation}
where, again, the dash denotes differentiation with respect to the scale factor \(a\).

We are most interested in the change in the radiation energy density $\Delta \rho_X$ as a consequence of dark matter--radiation coupling.  Expanding the energy density~(\ref{energydensitynu}) for small \(\xi_X \ll 1\) and $\Delta T_X/T_X \ll 1$, we find
\begin{eqnarray} 
\label{eq:changenu}
\frac{\Delta \rho_{\nu}}{\rho_{\nu}} &\simeq& 4 \frac{\Delta T_\nu}{T_\nu} + \frac{540\, \zeta(3)}{7\pi^{4}} \xi_{\nu},\\
 \label{eq:change}
\frac{\Delta \rho_{\gamma}}{\rho_{\gamma}} &\simeq & 4 \frac{\Delta T_\gamma}{T_\gamma} + \frac{90\, \zeta(3)}{\pi^{4}} \xi_{\gamma},
\end{eqnarray}
for neutrinos and photons respectively, where $\zeta(3) \simeq 1.202$ is the Riemann zeta function. 
Note that equation~(\ref{eq:changenu}) differs from the standard expression given in, e.g., equation (4.50) of reference~\cite{NuCosmology}, in that the former's correction from a non-zero chemical potential is linear in $\xi_\nu$ while the latter's is quadratic.  The difference stems from the standard assumption that  chemical potentials of equal magnitude but opposite signs for neutrinos and anti-neutrinos develop from a non-zero lepton asymmetry prior to the freeze-out of neutrino number-changing processes, e.g., $\nu \bar{\nu} \leftrightarrow e^+ e^-$.
 In contrast, our LKD scenario is lepton symmetric and the dark matter couples equally to neutrinos and anti-neutrinos: both populations, therefore, must develop identically the same chemical potential.
 
 This also means that our $\xi_\nu$  cannot be identified with the quantity probed by the light element abundances in the so-called degenerate big bang nucleosynthesis scenario~\cite{Mangano:2011ip}.  Constraints on $\xi_\nu$ from CMB anisotropy measurements~\cite{Oldengott:2017tzj} likewise do not apply.

\subsection{The coupled dark matter--radiation system}
\label{sec:master}

The central machinery for tracking the coupled dark matter--radiation system lies in the conservation of total comoving entropy in the coupled system, i.e., $\textrm{d}\left(s_Xa^{3} + s_{\psi}a^{3}\right) = 0$,
or equivalently from equations~(\ref{eq:DMentropy}) and~(\ref{eq:nuentropy}),
\begin{equation} \label{eq:ConEntropy}
\frac{\textrm{d}}{\textrm{d}a}\left\{N_X\left[4\frac{\textrm{Li}_{4}\left(\mp \textrm{e}^{\xi_{X}}\right)}{\textrm{Li}_{3}\left(\mp \textrm{e}^{\xi_X}\right)}-\xi_X\right] + N_{\psi}\left[|\xi_{\psi}|+x\frac{K_{1}(x)}{K_{2}(x)}\right]\right\} = 0,
\end{equation}
where we have defined \(N_{\psi} \equiv n_{\psi}a^{3}\) and \(N_X \equiv n_X a^{3}\) as the comoving dark matter and radiation number densities respectively; the former is always conserved, which may or may not be the case for the latter.
Then, enforcing the temperature relation $T_\psi = T_X$ and applying the dark matter number density conservation law~(\ref{eq:DMxi}), we obtain
 \begin{equation}
\begin{split} 
\label{eq:master}
&\left[3- 4 \frac{\textrm{Li}_{2}(\mp \textrm{e}^{\xi{_{X}}})
\textrm{Li}_{4}(\mp \textrm{e}^{\xi{_{X}}})}{\textrm{Li}_{3}^2(\mp \textrm{e}^{\xi_{X}})}\right] \frac{\xi_X'}{3}  + \frac{1}{3} \left[4\frac{\textrm{Li}_{4}\left(\mp \textrm{e}^{\xi_{X}}\right)}{\textrm{Li}_{3}\left(\mp \textrm{e}^{\xi_X}\right)}-\xi_X\right]  \frac{N_X'}{N_X}  \\
&\hspace{15mm}+ \frac{N_{\psi}}{N_X} \left\{ \frac{1}{a} + \left[1 - \frac{(2+x^{2})K^{2}_{1}(x)-x^{2}K^{2}_{0}(x) - x K_{0}(x)K_{1}(x)}{3 K^{2}_{2}(x)}\right] \frac{T_X^{'}}{T_X} \right\} = 0.
\end{split}
\end{equation}
This is the ``master equation'' of our analysis.


\section{Dark matter--neutrino elastic scattering}
 \label{Sec:Nu}
 
We now apply the formalism of section~\ref{sec:formalism} to the scenario of dark matter--neutrino elastic scattering.  This is an especially simple scenario in that after neutrino decoupling at $T \sim 1$~MeV, all neutrino and dark matter number-changing processes can be assumed to be absent.  Thus, comoving number density conservation applies equally to $\psi$ and $X=\nu$.


\subsection{Equations of motion}
\label{sec:nueom}

Setting $N_\nu'=0$ in the master equation~(\ref{eq:master})  and supplementing it with the number density  conservation law~(\ref{eq:nuxi}), we can now form two  equations of motion for the neutrino temperature~$T_\nu$ and degeneracy parameter $\xi_\nu$  respectively.  
 Parameterising the temperature as~\cite{Khatri:2011aj}, 
\begin{equation}
T_\nu(a) = T_\nu^{(0)}(a) \left[1+ \tau_\nu(a) \right],
\end{equation}
in which $T_\nu^{(0)}(a) \propto a^{-1}$ is the neutrino temperature  in the absence of dark matter--neutrino scattering,  we find
\begin{align}
\begin{split} \label{eq:NeutrinoDEs}
 \frac{\textrm{d} \ln (1+\tau_\nu)}{\textrm{d} \ln{a}}&= - \frac{b_1(x) (N_{\psi}/N_{\nu}) }{c_{1}(-y) + \left[1-b_{1}(x)\right](N_{\psi}/N_{\nu})} \simeq - \frac{b_{1}(x)}{c_{1}(-y)}(N_{\psi}/N_{\nu}), \\
\frac{\textrm{d}\xi_{\nu}}{\textrm{d} \ln{a}} &= \frac{3b_{2}(x) c_{2}(-y)(N_{\psi}/N_{\nu})}{c_{1} (-y)+ \left[1-b_{1}(x)\right](N_{\psi}/N_{\nu})} \simeq \frac{3b_{1}(x) c_{2}(-y)}{c_{1}(-y)}(N_{\psi}/N_{\nu}),
\end{split}
\end{align}
where the second equality in both expressions follows from $N_{\psi}/N_{\nu} \ll 1$, and 
\begin{equation}
\begin{aligned}
 \label{eq:b1}
b_{1}(x) &\equiv \frac{(2+x^{2})K^{2}_{1}(x)-x^{2}K^{2}_{0}(x) - x K_{0}(x)K_{1}(x)}{3 K^{2}_{2}(x)},\\
c_{1}(y) &\equiv 4 \frac{\textrm{Li}_{4}(y)}{\textrm{Li}_{3}(y)} - 3 \frac{\textrm{Li}_{3}(y)}{\textrm{Li}_{2}(y)}, \\
c_{2}(y) &\equiv  \frac{\textrm{Li}_{3}(y)}{\textrm{Li}_{2}(y)}, 
\end{aligned}
\end{equation}
 with $x \equiv m_\psi/T_\nu$, and $y \equiv e^{\xi_\nu}$ is the fugacity of the neutrino gas.
 
Observe here that $b_1\to 0$ as $x\to 0$ and $b_1\to 1/2$ as $x \to \infty$, so that the equations of motion~(\ref{eq:NeutrinoDEs}) ``switch on''  only
when the dark matter transitions from a relativistic to a non-relativistic species. Physically, this reflects the fact that energy exchange exclusively between relativistic particle species does not alter the relativistic energy density scaling from $\rho_\nu \propto a^{-4}$. The temperature scaling, therefore, remains \(T_\nu \propto a^{-1}\) until the dark matter becomes non-relativistic.


\subsection{Initial conditions and endpoints}

Any energy transfer from the neutrino bath to the dark matter sector occurring before the neutrinos have kinetically decoupled from the photon--electron--positron plasma will be communicated immediately to the latter. Consequently, changes to $N_{\rm eff}$ in LKD scenarios can only be meaningfully defined after neutrino decoupling, which we assume to have occurred instantaneously at $T_\nu(a_{\nu\textrm{dec}}) =1$~MeV, where \(a_{\nu\textrm{dec}} \simeq 1.7 \times 10^{-10}\).  
 
 We therefore solve equation~(\ref{eq:NeutrinoDEs}) beginning at  $a_{\nu \textrm{dec}}$  with the initial conditions
 \begin{equation}
\begin{aligned}
\label{eq:initial}
  T^{(0)}_\nu(a_{\nu {\rm dec}}) &=1~{\rm MeV}, \\
  \tau_\nu(a_{\nu {\rm dec}}) &= 0, \\
   \xi_\nu(a_{\nu {\rm dec}}) & =0,
 \end{aligned}
 \end{equation}  
  and integrate up to an endpoint $a_{\rm end}$ defined by
  \begin{equation}
  \label{eq:aend}  
    a_{\rm end} = {\rm min} \left(a_{\rm CMB}, a_{\psi {\rm  dec}} \right),
   \end{equation} 
where $a_{\rm CMB} \simeq 1 \times 10^{-4}$, and  $a_{\psi {\rm dec}}$ denotes the time at which the dark matter decouples from the neutrino bath. The former corresponds roughly to the time at which the smallest length scale probed by CMB anisotropy measurements enters the horizon, and  has been imposed here because changes to the relativistic energy density after this time affect the CMB anisotropies in ways that cannot be meaningfully quantified by a single $N_{\rm eff}$ parameter.

The latter, $a_{\psi {\rm dec}}$, can be determined from the decoupling condition \(\Gamma_{\textrm{relax}} (a_{\psi {\rm dec}})= H(a_{\psi {\rm dec}})\), where \(H\) is the Hubble expansion rate, and 
\begin{equation}
\label{eq:relax}
\Gamma_{\textrm{relax}} = \sqrt{\frac{3}{2}}\frac{T_X}{m_\psi} \sigma_{\psi-X}n_{X}
\end{equation}
is  the momentum relaxation rate~\cite{Hofmann:2001bi, Binder:2016pnr} given the DM--radiation scattering cross section $\sigma_{\psi-X}$, with $X=\nu$ in this instance.  Assuming a time-independent  $\sigma_{\psi-\nu}$ and that DM decoupling happens during radiation domination, the decoupling condition yields
\begin{equation}
a_{\psi {\rm dec }} \simeq  1.00 \times 10^{13} \left[\frac{\sigma_{\psi-\nu}}{m_{\psi}} \frac{\textrm{keV}}{\textrm{cm}^{2}}\right]^{1/2}
\end{equation}
as the DM decoupling time.

The fractional change in the neutrino energy density~(\ref{eq:changenu}) and hence in the effective number of neutrinos~(\ref{eq:deltaneff}) can then be computed from the solution by identifying $\Delta T_\nu/T_\nu \equiv \tau(a_{\rm end})$ and $\xi_\nu \equiv \xi_\nu(a_{\rm end})$.  


\subsection{Results}
\label{sec:nuresults}

\begin{figure}[t!]
	\begin{center}
		\includegraphics[width=0.8\textwidth]{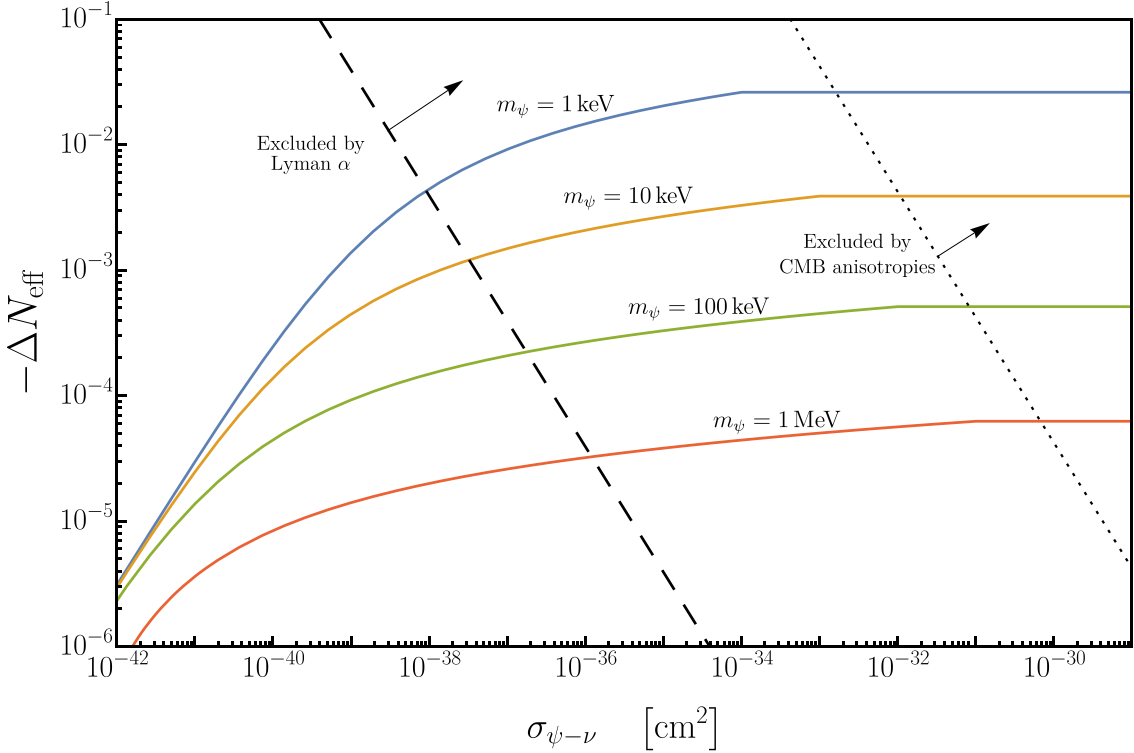}
		\caption{Absolute change in \(N_{\textrm{eff}}\) as a function of the dark matter--neutrino elastic scattering cross section $\sigma_{\psi-\nu}$ for different values of the DM mass \(m_{\psi}\). Here, the blue, orange, green, and red curves denote \(m_{\psi} = 1~\textrm{keV}, 10~\textrm{keV}, 100~\textrm{keV}\), and \(1~\textrm{MeV}\) respectively, while the dotted and dashed lines delineate the ($m_{\psi},\sigma_{\psi-\nu}$)-parameter regions excluded respectively by  CMB anisotropies and Lyman-\(\alpha\) measurements~\cite{Wilknu}. }
		\label{Fig:NuExact}
	\end{center}
\end{figure}

Figure~\ref{Fig:NuExact} shows the absolute change in  $N_{\rm eff}$  for a range of dark matter--neutrino scattering cross sections and dark matter masses, assuming the former to be time-independent.

In general,  the lighter the dark matter mass, the bigger the absolute value of the shift $|\Delta N_{\rm eff}|$.  This trend can be glimpsed already at the level of the equations of motion~(\ref{eq:NeutrinoDEs}), where the rates of change of $T_\nu$ and $\xi_\nu$ are directly proportional to the dark matter to neutrino number ratio $N_\psi/N_\nu$, which, for a fixed dark matter {\it energy density}, is inversely proportional to the DM mass.  Physically, this can be understood as a consequence of the neutrinos having to share more of their energy with a larger number of lighter mass DM particles in order to achieve equipartition of kinetic energy.
 
The subtle differences in the shapes of the $m_\psi$ curves can trace their origin to the dark matter's transition from a relativistic to a non-relativistic species in the timeframe of interest.  As discussed in section~\ref{sec:nueom}, the transfer of energy out of the neutrino population becomes efficient only when the dark matter becomes non-relativistic and the equations of motion~(\ref{eq:NeutrinoDEs}) ``switch on''.  In the limit $m_\psi = 0$ the energy transfer is identically zero.  Only for $m_\psi \gtrsim 1$~MeV is the dark matter 
non-relativistic for the entirety of the problem; the corresponding $m_\psi$ curves then resemble the approximate logarithmic solution considered in~\cite{Ali-Haimoud:2015pwa}.

In terms of the scattering cross section, larger values of $\sigma_{\psi - \nu}$ generally lead to larger $|\Delta N_{\rm eff}|$ shifts.  This is because, for a fixed DM mass, a larger cross section corresponds to a later DM kinetic decoupling time, and hence a longer period in which the dark matter can  siphon energy out of the neutrino population.  Note that all curves ``flatten out'' at  $\sigma_{\psi-\nu}/m_\psi\gtrsim 10^{-34}~{\rm cm}^{2} \ {\rm keV}^{-1}$  due to the artificial integration cut-off we have imposed at $a_{\rm CMB} \simeq 1 \times 10^{-4}$.  At the opposite end, for $\sigma_{\psi-\nu}/m_\psi \lesssim 3 \times 10^{-46}~{\rm cm}^{2}\ {\rm keV}^{-1}$, DM kinetic decoupling happens before neutrino decoupling, in which case  $\Delta N_{\rm eff} = 0$ by definition.


\section{Dark matter--photon elastic scattering} \label{Sec:Phot}

The case of dark matter--photon elastic scattering is complicated by the fact that photons interact in addition with the trace amounts of free electrons and nuclei---collectively, the baryons---that are present in the timeframe of interest.  In terms of their number densities, the baryon population represents a mere $\sim 10^{-9}$ of the photon population.  Nonetheless, interactions between these two sectors provide the photons with additional channels through which to lose their energy and even particle numbers.  These need to be taken into account in our modelling of the thermodynamics of the coupled dark matter--photon system.


\subsection{Equations of motion}

Roughly speaking, for photons, energy-changing processes such as Compton scattering can be taken to be operative for the entire period of interest, while number-changing processes such as double Compton scattering and Bremsstrahlung are efficient only at $a < a_{\mu} \simeq 5 \times 10^{-7}$~\cite{Chluba:2015bqa}.  In light of these ``background'' processes, we therefore divide our coupled dark matter--photon scenario into two regimes, labelled following the parlance of the CMB spectral distortion literature as:
\begin{enumerate}
\item The ``$T$-era'', defined as $a< a_\mu$, in which photons participate in both number- and energy-changing interactions with baryons, and
\item The ``$\mu$-era'', defined as $a > a_\mu$, where photons and baryons can only exchange energy with one another.
\end{enumerate}

\paragraph{The ${\bm T}$-era} 
Here, photon number-changing processes afforded by the baryons are able to drive the degeneracy parameter $\xi_\gamma$ to zero at all times. Thus, any energy gain or loss in the photon population as a consequence of new interactions can only manifest itself as a shift in the photon temperature $T_\gamma$, and hence the name ``$T$-era''.

Fixing  \(\xi_{\gamma} = 0\) and noting that the comoving photon number density is {\it not} conserved, i.e., $N_\gamma'\neq 0$, where  $N_{\gamma} \equiv n_{\gamma}(a) a^{3} = 2 \zeta(3) T_\gamma^3 a^3/\pi^2$, the master equation~(\ref{eq:master})  now reduces to a single equation of motion for the photon temperature~$T_\gamma$.  Parameterising $T_\gamma$ as~\cite{Khatri:2011aj}, 
\begin{equation}
T_\gamma(a) = T_\gamma^{(0)}(a) \left[1+ \tau_\gamma(a) \right],
\end{equation}
with $T_\gamma^{(0)}(a) \propto a^{-1}$,  we obtain
\begin{equation}
 \label{eq:DiffEq1}
\frac{\textrm{d} \ln{(1+\tau_\gamma)}}{\textrm{d} \ln{a}} = - \frac{b_1 (x) (N_{\psi}/N_{\gamma}) }{b_{0} + \left[1-b_{1}(x)\right](N_{\psi}/N_{\gamma})} \simeq  - \frac{b_{1}}{b_{0}}(N_{\psi}/N_{\gamma}),
\end{equation}
where the second equality follows from \(N_{\psi}/N_{\gamma} \ll 1\), $b_0$ is a constant given by
\begin{equation} 
\label{eq:FirstConsts}
b_{0} \equiv \frac{2\pi^{4}}{45\zeta(3)} \simeq 3.602, 
\end{equation}
and the function $b_1(x)$ can be found in equation~(\ref{eq:b1}).

Note that in deriving equation~(\ref{eq:DiffEq1}) we have neglected contributions from the baryons.  In general, we expect photon--baryon coupling to contribute a fractional correction to $N_{\rm eff}$ of order $N_b/N_\gamma \sim 10^{-9}$, an effect far too small to be probed by even the most idealised large-scale anisotropy surveys (see section~\ref{Sec:Forecast}) and which justifies our approximation. One caveat is the small window between neutrino decoupling at $T \sim 1$~MeV and $e^+e^-$ annihilation at $T \sim 0.5$~MeV, wherein the electron--positron and photon populations are comparable in size. The subsequent transfer of entropy during annihilation also alters the evolution of $T_\gamma^{(0)}$ substantially away from a simple $a^{-1}$ scaling.
  In practice, however, this window is much too narrow (in $\ln a$) to be of any appreciable effect on the final outcome.  We therefore leave equation~(\ref{eq:DiffEq1}) as is, and simply opt to initialise our calculation at $T \lesssim 0.5$~MeV, i.e., after $e^+ e^-$ annihilation, in order to circumvent these issues.%
\footnote{The dominant heating effect of the electron--positron population on the photon temperature relative to the neutrino temperature is already accounted for by the factor $(4/11)^{1/3}$ in equation~(\ref{eq:Defn}).}


\paragraph{$\bm \mu$-era} 
The $\mu$-era begins when photon number-changing processes drop out of equilibrium and only energy-changing processes remain efficient.  Here, the comoving photon number density is conserved, allowing a chemical potential to develop in the photon energy spectrum when energy is gained or lost.  This gives rise to the label ``$\mu$-era''. 

The situation is then exactly analogous to what we have already seen with dark matter--neutrino elastic scattering in section~\ref{Sec:Nu}, with the proviso that we again neglect contributions from the trace amounts of baryons.
  Thus, we can immediately generalise equation~(\ref{eq:NeutrinoDEs}) to
\begin{align}
\begin{split} \label{eq:PhotonDEs}
 \frac{\textrm{d} \ln (1+\tau_\gamma)}{\textrm{d} \ln{a}}&= - \frac{b_1(x) (N_{\psi}/N_{\gamma}) }{c_{1}(y) + \left[1-b_{1}(x)\right](N_{\psi}/N_{\gamma})} \simeq - \frac{b_{1}(x)}{c_{1}(y)}(N_{\psi}/N_{\gamma}), \\
\frac{\textrm{d}\xi_{\gamma}}{\textrm{d} \ln{a}} &= \frac{3b_{2}(x) c_{2}(y)(N_{\psi}/N_{\gamma})}{c_{1} (y)+ \left[1-b_{1}(x)\right](N_{\psi}/N_{\gamma})} \simeq \frac{3b_{1}(x) c_{2}(y)}{c_{1}(y)}(N_{\psi}/N_{\gamma}),
\end{split}
\end{align}
where $y \equiv e^{\xi_\gamma}$ is the fugacity of the photon gas, and the functions  $b_1(x), c_1(y), c_2(y)$ are given in equation~(\ref{eq:b1}).


\subsection{Initial conditions and endpoints} 
\label{sec:Solns}

We solve equation~(\ref{eq:DiffEq1}) at $a < a_\mu \simeq 5 \times 10^{-7}$ and equation~(\ref{eq:PhotonDEs}) at $a > a_\mu$ for the photon temperature perturbation $\tau_\gamma(a)$ and degeneracy parameter~$\xi_\gamma(a)$. Note that this abrupt switch between sets of equations at \(a \approx a_{\mu}\) is valid as \(\tau_{\gamma}(a)\) remains continuous and free of unphysical `kinks.' As discussed above, we initialise at $e^+e^-$ annihilation, assumed to have occurred instantaneously at  $T_\gamma(a_{\rm ann}) = 0.5$~MeV, where $a_{\rm ann} \simeq 4.7 \times 10^{-10}$, with the  initial conditions
\begin{equation}
\begin{aligned}
\label{eq:initialgamma}
  T^{(0)}_\gamma(a_{\rm ann}) &=0.5~{\rm MeV}, \\
  \tau_\gamma(a_{\rm ann}) &= 0, \\
   \xi_\gamma(a_{\rm ann}) & =0.
 \end{aligned}
 \end{equation}  
The integration is performed up to an endpoint $a_{\rm end}$  formally defined in equation~(\ref{eq:aend}).

The dark matter--photon decoupling time is quantified again by the decoupling condition \(\Gamma_{\textrm{relax}} (a_{\psi {\rm dec}})= H(a_{{\psi \rm dec}})\), where the momentum relaxation rate $\Gamma_{\rm relax}$ is likewise given formally by equation~(\ref{eq:relax}), with $X=\gamma$.
Assuming a time-independent cross section  $\sigma_{\psi-\gamma}$ and that DM decoupling happens during radiation domination, the decoupling time evaluates to
\begin{equation}
a_{\psi {\rm dec }} \simeq  1.30 \times 10^{13} \left[\frac{\sigma_{\psi-\gamma}}{m_{\psi}} \frac{\textrm{keV}}{\textrm{cm}^{2}}\right]^{1/2},
\end{equation}
from which the integration endpoint $a_{\rm end}$ can be determied via equation~(\ref{eq:aend}).

Identifying $\Delta T_\nu/T_\nu \equiv \tau(a_{\rm end})$ and $\xi_\nu \equiv \xi_\nu(a_{\rm end})$,  the fractional change in the photon energy density and hence in the effective number of neutrinos can then be computed from the solution via equations~(\ref{eq:change}) and~(\ref{eq:deltaneff}).

    
\subsection{Results} \label{Sec:Results}

\begin{figure}[t!]
	\begin{center}
		\includegraphics[width=0.8\textwidth]{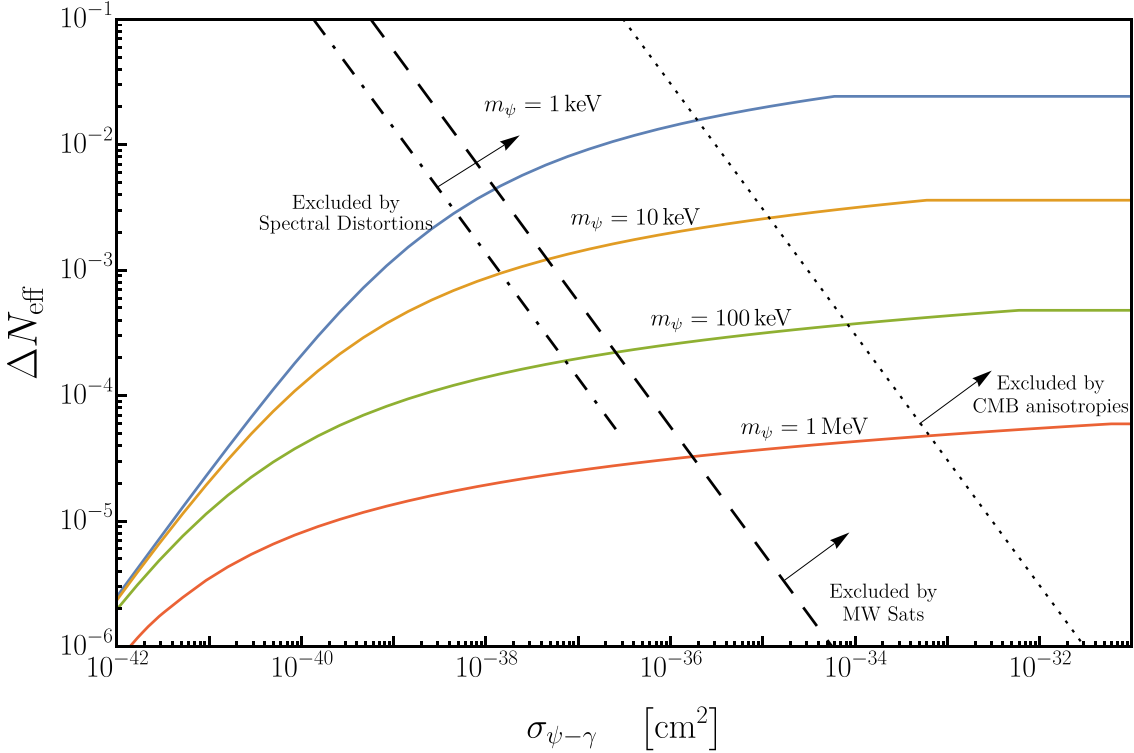}
		\caption{Absolute change in \(N_{\textrm{eff}}\) as a function of the dark matter--photon elastic scattering cross section for different values of the DM mass \(m_\psi\). Here, the blue, orange,  green, and curves denote \(m_{\psi} = 1~\textrm{keV}, 10~\textrm{keV}, 100~\textrm{keV}\) and \(1~\textrm{MeV}\) respectively. The dotted, dashed, and dot-dashed lines delineate the ($m_{\psi}$-$\sigma_{\psi-\gamma}$)-parameter regions excluded respectively  by CMB anisotropies~\cite{Wilkphoton}, Milky Way satellite counts~\cite{Boehm:2014vja}, and CMB spectral distortions~\cite{Ali-Haimoud:2015pwa}.}		\label{Fig:PhotonExact}
	\end{center}
\end{figure}

Figure~\ref{Fig:PhotonExact} shows the absolute change in  $N_{\rm eff}$  for a range of dark matter--photon scattering cross sections and dark matter masses, assuming the former to be time-independent.  Apart from the sign flip, qualitatively these results are very similar to those for dark matter--neutrino coupling presented in figure~\ref{Fig:NuExact} and discussed in detail in section~\ref{sec:nuresults}.

Of particular note in figure~\ref{Fig:PhotonExact} is the region $\sigma_{\psi-\gamma}/m_\psi\lesssim 1.5 \times 10^{-39}~{\rm cm}^{2} \ {\rm keV}^{-1}$, i.e., to the left of the dot-dashed line labelled ``Excluded by spectral distortions''.  This region corresponds to photon energy loss exclusively in the $T$-era, which manifests solely as a photon temperature shift and hence cannot be observed as a distortion to the blackbody spectrum of the CMB photons~\cite{Ali-Haimoud:2015pwa}.  Our results here indicate that the energy loss can instead be observed as a small shift in the $N_{\rm eff}$ parameter.


\section{Parameter sensitivity forecast} \label{Sec:Forecast}

We have found in sections~\ref{Sec:Nu} and~\ref{Sec:Phot} that  the change in \(N_{\textrm{eff}}\) due to elastic scattering to be within the reach of the next generation of cosmological probes such as CMB-S4 only for extremely light dark matter masses, e.g., $m_\psi = 1$~keV produces a maximum shift of \(\Delta N_{\textrm{eff}} \simeq -0.027\) and \(\Delta N_{\rm eff} \simeq 0.013\) respectively for DM--neutrino and DM--photon coupling.

However,  large swaths of the parameter regions responsible for these optimistic values are already excluded by current CMB anisotropy and, in the case of DM--$\gamma$ coupling, by spectral distortion bounds on $\sigma_{\psi-X}/m_\psi$~\cite{Wilknu,Wilkphoton,Ali-Haimoud:2015pwa}.
Small-scale nonlinear observables such the Lyman-$\alpha$ forest and Milky Way satellite counts likewise constrain $\sigma_{\psi-X}/m_\psi$ via its role in collisional damping of the dark matter perturbations~\cite{Wilknu,Boehm:2014vja}.  These same observables---especially the Lyman-$\alpha$ forest---also limit the dark matter particle mass to $m_\psi \gtrsim O({\rm keV})$ in order to circumvent excessive free-streaming damping.  While bounds derived from nonlinear observables need to be interpreted with care, realistically we expect the as-yet-unprobed parameter regions to generate $N_{\rm eff}$ shifts of no more than  $|\Delta N_{\rm eff}| \sim O(10^{-3})$.

It is then interesting to ask whether these parameter regions could be explored at all via their effect on $N_{\rm eff}$ by any cosmological observation.  To this end, we consider a hypothetical large-scale structure tomographic high-redshift survey limited only by cosmic variance.   Such a survey might be, e.g.,  a futuristic 21~cm tomography measurement that can access the neutral hydrogen spin-flip signal during the ``dark ages'' $30 \lesssim z \lesssim 300$~\cite{Loeb:2003ya}.   We shall conduct a parameter sensitivity forecast to determine how well such an observable can constrain $N_{\rm eff}$.

 
 \subsection{Fisher matrix}
 
 Following~\cite{Chen:2016zuu}, we assume for definiteness a 21~cm tomography measurement of the 3D matter power spectrum $P(k,z)$ in 
 the redshift range \(30 \leq z \leq z_{\rm max}\), divided into $n_z$~bins of equal widths~\(\Delta z = 5\), where  the maximum redshift $z_{\rm max}$ may vary from 40 to 100.  In the absence of shot noise,
the constraining power of such an idealised survey in relation to cosmological parameters~$\theta_\alpha$ can be assessed using the Fisher information matrix,
\begin{equation} 
\label{eq:Fisher}
F^{\rm 21cm}_{\alpha \beta} = \sum_i^{n_z} V_i \int^{k_{\textrm{max}}}_{k_{\textrm{min},i}} \, \frac{k^{2} \, \textrm{d}k}{2\pi^{2}} \frac{\partial P(k,z_i)}{\partial \theta_{\alpha}} \frac{1}{P^2(k,z_i)} \frac{\partial P(k,z_i)}{\partial \theta_{\beta}},
\end{equation}
where 
\begin{equation}
V_i = \frac{4\pi}{3}\left[d(z_{\textrm{max},i})^{3}-d(z_{\textrm{min},i})^{3}\right]
\end{equation}
is the comoving volume of the $i$th redshift bin at $z \in [z_{{\rm low},i},z_{{\rm high},i}]$, and 
\begin{equation}
d(z) = c\int \, \frac{\textrm{d}z'}{H(z')}
\end{equation}
is the comoving distance at $z$.  The minimum wavenumber accessible in each redshift bin is set  by \(k_{\textrm{min},i} = 2\pi\left(3V_i/4\pi\right)^{-1/3}\); we use however a common $k_{\rm max}$ in all bins, up to an absolute maximum of \(k_{\max} = 300~\textrm{Mpc}^{-1}\), roughly the baryon Jeans scale.  For simplicity we do not model the bias between the brightness temperature and the matter density fluctuations.

We vary six parameters in our computation of $F^{\rm 21cm}_{\alpha \beta}$, namely,
\begin{equation}
 \theta_{\alpha} \in \left( \omega_{b}, \omega_{\psi}, A_{s}, n_{s}, H_{0}, N_{\textrm{eff}} \right),
\end{equation}
where  $\omega_{b}$ is the physical baryon density, $\omega_{\psi}$ the physical dark matter density, $H_0$ the present-day Hubble parameter, $A_{s}$ the amplitude of the primordial curvature power spectrum, and  $n_{s}$ its spectral index.  Note that we have omitted the canonical~\(\tau_{\textrm{reio}}\), the optical depth to reionisation, because the 3D matter power spectrum is not sensitive to this parameter.  We evaluate the derivatives of $P(k,z)$ respect to $\theta_\alpha$ at $\theta_\alpha = \theta_\alpha^{\rm fid}$ by finite differencing, where the fiducial parameter values~$\theta_\alpha^{\rm fid}$ are set by the 
 Planck 2018 best-fit~\cite{Aghanim:2018eyx}, and we  compute the power spectra $P(k,z)$ using the Boltzmann code~\texttt{CLASS}~ \cite{CLASSI,CLASSII}.%
 \footnote{Available at \href{http://class-code.net/}{http://class-code.net/}}

The information content of this hypothetical 21~cm survey should be assessed together with what could be achieved by a future CMB measurement, as it is generally understood that parameters such as the physical baryon density~$\omega_b$ can be very tightly constrained by the latter class of probes.  We take for concreteness CMB-S4, and construct the corresponding (inverse) Fisher matrix from the $1 \sigma$-sensitivities $\sigma(\theta_\alpha)$ given in~\cite{Abazajian:2016yjj},
 \begin{equation}
 (F^{\textrm{CMB}})^{-1}_{\alpha \beta} = \delta_{\alpha \beta} \, \sigma^2(\theta_{\alpha}), 
 \end{equation}
 where $\delta_{\alpha \beta}$ is a Kronecker delta.
 The total information content is then quantified by the sum
\begin{equation}
\label{eq:totalfisher}
F^{\textrm{total}}_{\alpha \beta} = F^{\rm 21cm}_{\alpha \beta} + F^{\textrm{CMB}}_{\alpha \beta},
\end{equation}
 assuming $F^{\rm 21cm}_{\alpha \beta}$ and  $F^{\textrm{CMB}}_{\alpha \beta}$ to be uncorrelated,  which is well justified as the primary signals of the two probes originate from density fluctuations in mutually exclusive regions of space.


\subsection{Sensitivity to $N_{\rm eff}$}


\begin{figure}[t!]
	\begin{center}
		\includegraphics[width=0.40\textwidth]{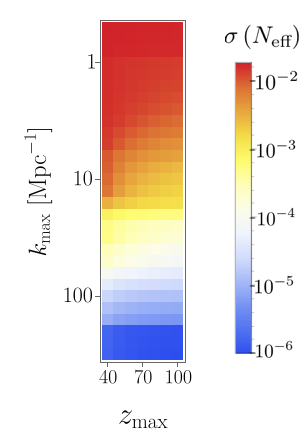}
		\caption{Forecasted $1\sigma$ sensitivity to $N_{\rm eff}$, \(\sigma \left(N_{\textrm{eff}}\right)\), for a cosmic variance-limited \(21~\textrm{cm}\) tomography survey at  $30 \lesssim z \lesssim z_{\rm max}$ combined with projected bounds from future CMB-S4 data~\cite{Abazajian:2016yjj}, as a function of the wavenumber and redshift cut-offs, $k_{\rm max}$ and $z_{\rm max}$.}
				\label{Fig:Fisher}
	\end{center}
\end{figure} 


The projected $1 \sigma$ sensitivity to the $\alpha^{\textrm{th}}$ parameter can now be computed from the total Fisher matrix~(\ref{eq:totalfisher}) as
\begin{equation}
\sigma(\theta_{\alpha}) = \sqrt{\left(F^{\textrm{total}}\right)^{-1}_{\alpha \alpha}}.
\end{equation}
We explore a range of maximum wavenumbers~$k_{\rm max}$ and maximum redshifts $z_{\rm max}$ for our hypothetical  21~cm tomography survey.
Figure~\ref{Fig:Fisher} shows the possible sensitivities \(\sigma\left(N_{\textrm{eff}}\right)\) that could be achieved for different combinations of these experimental settings.

Clearly, in terms of $k_{\rm max}$, figure~\ref{Fig:Fisher} shows that one needs to go significant beyond $k_{\rm max} \sim 1~{\rm Mpc}^{-1}$ in order to improve on  CMB-S4's projected sensitivity to $N_{\rm eff}$ of  $\sim 0.02$.
This is perhaps not surprising, since CMB-S4 derives its constraining power from $B$-mode lensing measurements of the large-scale structure, which contains some of the same information as our hypothetical 21~cm survey.
Extremely optimistic configurations, e.g., \(k_{\textrm{max}} = 300~\textrm{Mpc}^{-1}$ and $z_{\textrm{max}} = 100\), could yield \(\sigma \left(N_{\textrm{eff}}\right) \simeq 10^{-6}\), enabling the entire parameter space of figures \ref{Fig:NuExact} and \ref{Fig:PhotonExact} to be probed. For more conservative choices, e.g., \(k_{\textrm{max}} = 25~\textrm{Mpc}^{-1}$ and $z_{\textrm{max}} = 40\), we find \(\sigma \left(N_{\textrm{eff}}\right) \simeq 5 \times 10^{-4}\), which still offers a window to  DM masses of order \(10\to100~\textrm{keV}\).


\section{Conclusions} \label{Sec:Conclusion}

In this work, we have computed the change in the effective number of neutrinos \(N_{\textrm{eff}}\) due to the hypothetical elastic scattering of dark matter particles with photons or neutrinos, motivated by late kinetic decoupling scenarios that may be of relevance to the small-scale crisis of dark matter. Assuming equilibrium thermodynamics, we modelled the drainage of kinetic energy from the radiation to the dark matter sector, and in particular the transition of the dark matter from a relativistic to a non-relativistic species which kicks off the drainage process.

In both cases of dark matter--neutrino and dark matter--photon coupling, we find the shift \(|\Delta N_{\textrm{eff}}|\) to increase with the scattering cross section~$\sigma_{\psi-X}$ but to scale inversely with the dark matter particle mass~$m_\psi$.  For $(m_\psi,\sigma_{\psi-X})$-parameter regions already explored by nonlinear probes such as the Lyman-$\alpha$ forest through collisional and/or free-streaming damping of the small-scale density fluctuations, we find shifts of $|\Delta N_{\rm eff}| \simeq O(10^{-2})$, which may be within the reach of future CMB anisotropy measurements such as CMB-S4.  For the bulk of the as-yet-unprobed parameter space, however, we generally expect $|\Delta N_{\rm eff}| \lesssim O(10^{-3})$.

This then prompted us to perform a  parameter sensitivity forecast for a hypothetical \(21~\textrm{cm}\) tomography survey of the large-scale structure in the dark ages,  in order to determine how much of the $(m_\psi,\sigma_{\psi-X})$-parameter space could be probed in theory. We find that an ideal survey limited only by cosmic variance up to $k_{\rm max} \simeq 300~{\rm Mpc}^{-1}$ in the redshift range $30 \lesssim z \lesssim 100$ is potentially sensitive to $|\Delta N_{\rm eff}| \simeq O(10^{-6})$.  If such a survey were ever realised, sensitivity of this order would enable us to probe dark matter masses up to \(m_{\psi} \sim 10~\textrm{MeV}\).


\acknowledgments

JADD acknowledges support from an Australian Government Research
Training Program Scholarship. The work of Y$^3$W is supported in part by the Australian
Government through the Australian Research Council’s Discovery Projects funding scheme
(project DP170102382).


\bibliography{ref}

\bibliographystyle{utcaps}

\end{document}